\documentclass[12pt]{iopart}
\usepackage{url}
\usepackage{graphicx}
\usepackage[dvipdfmx]{hyperref}
\usepackage{color}
\begin{document}

\title{Cryogenic suspension design for a kilometer-scale gravitational-wave detector}

\author{Takafumi Ushiba${}^{1}$, Tomotada Akutsu${}^{2}$, Sakae Araki${}^3$, Rishabh Bajpai${}^4$, Dan Chen${}^{2}$, Kieran Craig${}^1$, Yutaro Enomoto${}^{5}$, Ayako Hagiwara${}^1$, Sadakazu Haino${}^{6}$, Yuki Inoue${}^{3,7,8}$, Kiwamu Izumi${}^9$, Nobuhiro Kimura${}^3$, Rahul Kumar${}^3$, Yuta Michimura${}^{10}$, Shinji Miyoki${}^1$, Iwao Murakami${}^3$, Yoshikazu Namai${}^3$, Masayuki Nakano${}^{11}$, Masatake Ohashi${}^1$, Koki Okutomi${}^{12}$, Takaharu Shishido${}^4$, Ayaka Shoda${}^{2}$, Kentaro Somiya${}^{13}$, Toshikazu Suzuki${}^{1}$, Suguru Takada${}^{14}$, Masahiro Takahashi${}^1$, Ryutaro Takahashi${}^{2}$, Shinichi Terashima${}^3$, Takayuki Tomaru${}^{2,3}$, Flavio Travasso${}^{15,16}$, Ayako Ueda${}^3$, Helios Vocca${}^{16,17}$, Tomohiro Yamada${}^{10}$, Kazuhiro Yamamoto${}^{11}$, and Simon Zeidler${}^{18}$}

\address{${}^1$Institute for Cosmic Ray Research, The University of Tokyo, Kashiwa, Chiba 277-8582, Japan\\
${}^2$National Astronomical Observatory of Japan, Mitaka, Tokyo, 181-8588, Japan\\
${}^3$High Energy Accelerator Research Organization (KEK), Tsukuba, Ibaraki 305-0801, Japan\\
${}^4$The Graduate University for Advanced Studies (SOKENDAI), Tsukuba, Ibaraki 305-0801, Japan\\
${}^5$Department of Applied Physics, University of Tokyo, Bunkyo, Tokyo 113-8656, Japan\\
${}^6$Institute of Physics, Academia Sinica, Nankang, Taipei 11529, Taiwan\\
${}^7$Department of Physics, National Central University, No. 300, Zhongda Rd., Zhongli District, Taoyuan, 32001, Taiwan\\
${}^8$Center for High Energy and High Field Physics, National Central University, No. 300, Zhongda Rd., Zhongli District, Taoyuan, 32001, Taiwan\\
${}^9$Institute of Space and Astronautical Science, Japan Aerospace Exploration Agency, 3-1-1 Yoshinodai, Chuo-ku, Sagamihara, Kanagawa, 252-5210, Japan\\
${}^{10}$Department of Physics, The University of Tokyo, Bunkyo, Tokyo 113-0033, Japan\\
${}^{11}$Department of Physics, University of Toyama, Toyama, Toyama 930-8555, Japan\\
${}^{12}$The Graduate University for Advanced Studies (SOKENDAI), Mitaka, Tokyo 181-8588, Japan\\
${}^{13}$Department of Physics, Tokyo Institute of Technology, Meguro, Tokyo 152-8551, Japan\\
${}^{14}$National Institute for Fusion Science, Toki, Gifu, 509-5292, Japan\\
${}^{15}$School of Sciences and Technology, University of Camerino, 62032 Camerino, Italy\\
${}^{16}$The National Institute for Nuclear Physics (INFN), sez. di Perugia, 06123 Perugia, Italy\\
${}^{17}$Department of Physics and Geology, The University of Perugia, 06123 Perugia, Italy\\
${}^{18}$Department of Physics, Rikkyo University, 3-34-1 Nishi-Ikebukuro, Toshima-ku, Tokyo 171-8501, Japan}
\ead{ushiba@icrr.u-tokyo.ac.jp}
\vspace{10pt}
\begin{indented}
\item[]25 January 2021
\end{indented}

\begin{abstract}
We report the mirror suspension design for Large-scale Cryogenic Gravitational wave Telescope, KAGRA, during bKAGRA Phase 1. Mirror thermal noise is one of the fundamental noises for room-temperature gravitational-wave detectors such as Advanced LIGO and Advanced Virgo. Thus, reduction of thermal noise is required for further improvement of their sensitivity. One effective approach for reducing thermal noise is to cool the mirrors. There are many technical challenges that must be overcome to cool the mirrors, such as cryocooler induced vibrations, thermal drift in suspensions, and reduction in duty cycling due to the increased number of potential failure mechanisms.
Our mirror suspension has a black coating that makes radiative cooling more efficient. For conduction cooling, we developed ultra high purity aluminum heat links, which yield high thermal conductivity while keeping the spring constant sufficiently small. A unique inclination adjustment system, called moving mass, is used for aligning the mirror orientation in pitch. Photo-reflective displacement sensors, which have a large range, are installed for damping control on marionette recoil mass and intermediate recoil mass. Samarium cobalt magnets are used for coil-magnet actuators to prevent significant change of magnetism between room temperature and cryogenic temperature. In this paper, the design of our first cryogenic payload and its performance during bKAGRA Phase 1 are discussed.
\end{abstract}

\section{Introduction}%%%%%%%%%%%%%%%%%%%%%%%%%%%%%%%%%%%%%%%%%%%%%%%%%%%%%%%%%%%%%%%%%%%%%%%%%%%%%%%%%%%%%%%%%%%%%%%%%%%%%%
In 2015, Advanced LIGO detected gravitational waves (GWs) from a binary black hole coalescence \cite{GW150914}. Since then, many black hole mergers and a neutron star binary coalescence have been detected \cite{GWTC-1}. These GW observations enabled the test of general relativity under strong and dynamic gravity \cite{GRT150914}, rate estimates of compact binary coalescences \cite{BHBHCRE1,NSNSCRE1}, revealed parameters of black holes which indicates the fact that the black hole masses appear to be outside the previously observed ranges \cite{BBHMO1}, and masure of the Hubble constant \cite{HCM170817}. Thus, they opened an era of GW physics and astronomy. For further development of the field, it is important that detector sensitivity improves so that more GW events are detected with higher fidelity.

The ground-based GW detectors such as LIGO, Virgo, and KAGRA have to detect tiny displacements on the order of $10^{-19}\,\rm{m}$ of the mirrors that are placed several kilometers apart.  In order to reach such a high sensitivity, it is necessary to reduce displacement noise induced by the environmental effects as much as possible. Materials at a finite temperature cause effective displacement called thermal noise. Since the mirror thermal noise is one of the dominant noise sources of existing laser interferometric GW detectors at room temperature \cite{LIGO, Virgo}, it is important to reduce mirror thermal noise in GW detectors for improving their sensitivities. One effective way to reduce mirror thermal noise is to cool a mirror at cryogenic temperature.

KAGRA \cite{KAGRA} is an interferometric GW detector located in Japan. KAGRA has two unique features: it is constructed underground as seismic motion is smaller, and it uses cryogenic mirrors (20\,K) as test masses to reduce thermal noise. In addition, thermal noise can also be reduced by utilizing material with high mechanical quality factor (Q-factor) for the mirror and its suspension, KAGRA uses sapphire, which have a high mechanical Q-factor close to $10^8$ at cryogenic temperatures \cite{sapphireQ}, for mirrors and their suspension fibers. However, there are several technical difficulties such as heat extraction from mirrors, reduction of vibration induced from cryocoolers, and mitigation of suspension drift due to thermal contraction.

The first cryogenic mirror suspension of KAGRA was installed on 30th of November, 2017. Subsequently, it was cooled down from February and achieved a temperature below 20\,K in March, 2018. Then, a part of characterization of the cryogenic mirror suspension was done during a test operation called bKAGRA Phase 1 \cite{bKAGRA-phase1}. In this paper, we summarize our suspension design and its performance.

\section{Mechanical design of a cryogenic payload}%%%%%%%%%%%%%%%%%%%%%%%%%%%%%%%%%%%%%%%%%%%%%%%%%%%%%%%%%%%%%%%%%%%%%%%%%%%%%
\subsection{Overview}
\begin{figure}
\begin{center}
\includegraphics[width=10cm]{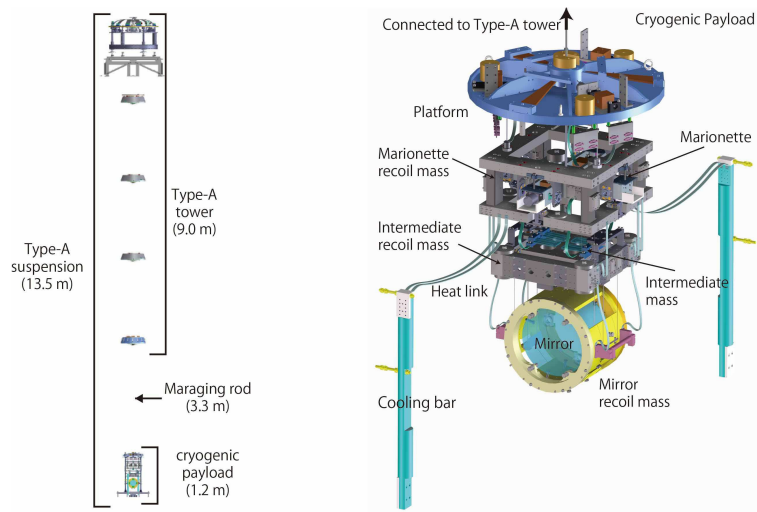}
\end{center}
\caption{(Left) Appearance of Type-A tower. Type-A tower consists of five stages, F0, F1, F2, F3, and BF from the top. F0 is set on the inverted pendulum for low-frequency horizontal vibration isolation. The figure is adapted and revised from Ref.\cite{bKAGRA-phase1} (Right) Appearance of a cryogenic payload. This figure is also adapted and revised from Ref.\cite{bKAGRA-phase1}. Heat links are connected between the marionette recoil mass and cooling bars, which are connected thermally well with the 2nd stage of cryocoolers.}
\label{Type-A}
\end{figure}

The KAGRA test masses (sapphire mirrors) are suspended by a nine-stage vibration isolation system, 13.5\,m in height, named Type-A suspension. The upper five stages, collectively called Type-A tower, are at room temperature and the lower four stages, the cryogenic payload, are cooled down to 20\,K. The four stages are platform (PF), marionette (MN), intermediate mass (IM), and test mass (TM) from the top. The TM chain, which consists of TM, IM, and MN, are surrounded by their corresponding recoil masses (RMs) in which displacement sensors and actuators \cite{MAP} are mounted. The TM is made of sapphire and the RM is made of low-magnetism stainless steel so as not to induce unexpected magnetic force on the suspension. The other suspensions are made of stainless steel of SUS316L. Figure \ref{Type-A} shows drawings of the Type-A tower and the cryogenic payload.

PF is suspended from the bottom stage of the Type-A tower (Bottom filter, BF) by a 3.3\,m-long Maraging steel wire. A TM chain and corresponding RMs (RM chain) are independently suspended from the PF. The TM chain is suspended from the PF by a single Maraging steel rod hung off 3 copper beryllium, (CuBe) blade springs.  IM is suspended from MN by four CuBe wires and TM is suspended from IM by 4 sapphire fibers hung on 4 sapphire blade springs on the IM. The RM chain has a similar suspension with three CuBe wires, four CuBe wires, and two looped CuBe wires.

\begin{figure}
\begin{center}
\includegraphics[width = 14cm]{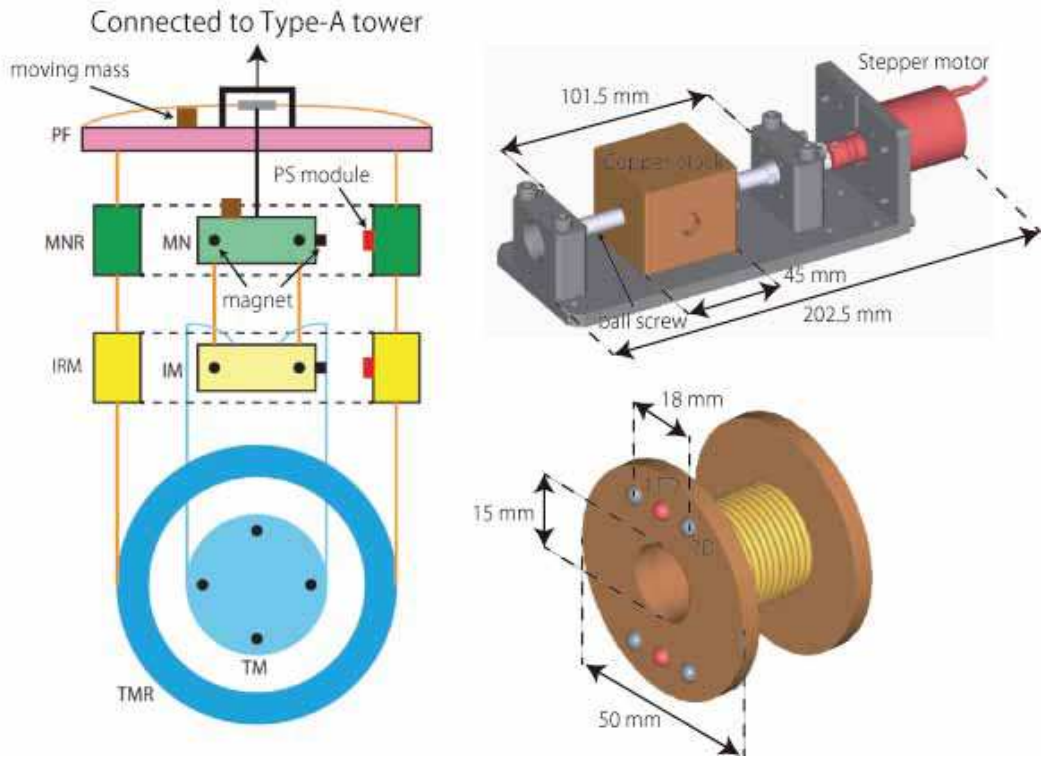}
\end{center}
\caption{(Left) Schematic view of the cryogenic payload. Black dots and squares on a TM chain represent magnets, and red squares on a RM chain show a photosensor module. Brown squares on PF and MN represent moving masses. (Right top) Schematic view of moving mass system. Red part is a stepper motor and silver part is a ball screw. Brown and gray parts are copper block and fixture, respectively. (Right bottom) Schematic view of a PS module. Gold lines are coil wires. Red and gray hemispheres represent LEDs and PDs, respectively. PDs are set on the both side of LEDs to reduce coupling between translational and tilt motion.}
\label{Fig1}
\end{figure}

Since we utilize a geometric anti-spring (GAS) filter \cite{GAS} for vertical vibration isolation in a room temperature suspension, the total weight of a cryogenic payload must be adjusted to properly load of the GAS, which has a design load of 200\,kg. For this reason, the total weight of the cryogenic payload design weight is as 198\,kg to maintain a safety margin, such that additional weights can be added to finely tune the GAS load. Table \ref{Tab1} is a summary of each stage of the cryogenic payload. Figure \ref{Fig1} shows a schematic view of configuration of the cryogenic payload. Using these parameters, a rigid body model simulator named SUMCON was developed \cite{SUMCON}. From this model and typical seismic spectrum at KAGRA site, seismic coupling has been estimated, which is well below KAGRA requirement \cite{MAP}.
\begin{table}
\caption{Summary of the mechanical design of each stage. Both test mass chain and recoil mass chain are suspended from platform (PF) separately. Test mass stage consists of marionette (MN), intermediate mass (IM), and test mass (TM) from the top. recoil mass chain also consists of three stages: marionette recoil mass (MNR), intermediate recoil mass (IRM), and recoil mass (RM) from the top.}
\label{Tab1}
\begin{center}
\begin{tabular}{cccccc}
Stage & Weight & Suspension wire & Number of wire & Wire length & Wire thickness\\\hline
PF & 65\,kg & Maraging steel & 1 & 3300\,mm & 12\,mm\\
MN & 21\,kg & Maraging steel & 1 & 345\,mm & 7\,mm \\
IM & 19\,kg & Copper Beryllium & 4 & 241.3\,mm & 0.6\,mm \\
TM & 23\,kg & Sapphire & 4 & 350\,mm & 1.6\,mm \\
MNR & 21\,kg & Copper Beryllium & 3 & 138\,mm & 7\,mm \\
IRM & 20\,kg & Copper Beryllium & 4 & 242.7\,mm & 0.6\,mm \\
RM & 29\,kg & Copper Beryllium & 2 & 243.5\,mm & 0.6\,mm \\\hline
\end{tabular}
\end{center}
\end{table}

There are three kinds of local sensors on a cryogenic payload for local control of suspensions. One is angular sensing optical levers (AS OpLevs) on MN and TM for sensing their angular motions \cite{oplev}. Another is a length sensing optical lever (LS OpLev) on TM for sensing its motion along optical axis of the main interferometer \cite{oplev}. The other is photo-reflective displacement sensors (PSs), which consists of one light-emitting diode (LED) and two photo detectors (PDs), on marionette recoil mass (MNR) and intermediate recoil mass (IRM) for sensing relative displacement between a TM chain and an RM chain. LED emits light and it is reflected on the well-polished stainless steel surface of MN and IM. Since the amount of detected light on PD, which is set on the same surface of LED, is changed in distance between the PS and reflecting surface, relative displacement between the TM chain and the RM chain can be measured. Room-temperature suspensions of LIGO and KAGRA use shadow sensors \cite{OSEM,OSEMK}, which have several mm range. However, these are not appropriate for KAGRA. Since each suspension wire has slightly different coefficient of thermal expansion due to the individual difference, each wire length varies to different lengths and causes misalignment when cooling. In addition, suspension wires are twisted when they shrink due to thermal contraction. Since misalignment due to the thermal contraction can be in order of mrad, larger range sensors are required for KAGRA. The photo-reflective sensors have a range of 10\,mm, adequate to accommodate KAGRA thermal contraction. Details of PS will be given in a separate paper \cite{PS}.

There are two kinds of actuators on the cryogenic payload for the local and global control of suspensions. The first actuator is a coarse alignment control actuator called {\it moving mass} on PF and MN. Figure \ref{Fig1} shows a schematic view of the moving mass system. The moving mass changes the center of mass of PF and MN, this results in inclination control on the RM chain and TM chain, respectively. The moving mass system consists of three components: a ball screw, copper block, and stepper motor. The ball screw is an oil-free type to be compatible with ultra-high vacuum and a lubricated system would likely freeze making the actuator stick at cryogenic temperatures. The copper block is a simple rectangular box shape, which is 45\,mm in length, 45\,mm in width, and 40\,mm in height.  The block has a metal spherical base for low friction. The stepper motor is a cryogenic compatible model. The whole system is 202.5\,mm in length, and actuation range and resolution of moving mass is $\pm$ 24.4\,mm and 1.25\,${\rm \mu m}$, respectively. The second type of actuator is a coil-magnet actuator that is utilized for precise control of the suspension. MN, IM, and TM have magnets and their RMs have corresponding coils. We can provide force to the suspensions by applying current to the coils. MN and IM stage have 6 actuators for each to actuate 6 degrees of freedom. TM stage has 4 actuators to actuate 3 degrees of freedom: longitudinal, yaw and pitch. All magnets are samarium-cobalt (SmCo), this magnet material is chosen as its magnetism does not change significantly between room temperature and cryogenic temperatures \cite{magnet}. Each PS sensor is combined with a coil actuator to form a compact PS module that provides sensing and actuation in one degree of freedom. Figure \ref{Fig1} shows a schematic view of a PS module.

\subsection{Platform stage}
The separate suspension of the TM chain and the RM chain allows reduced coupling of vibration from cryocoolers to the TM chain. This is accomplished by attaching the heat link to the MNR such that the PF provides lateral isolation. Moreover, PF has three CuBe blade springs, which have bounce modes with resonant frequency of 3.5\,Hz. Therefore, the PF stage works as not only lateral but also vertical isolator. The CuBe blade springs are installed with an angle of 7.5 degrees in order for their tips to be flat when the TM chain is suspended.

\begin{figure}
\begin{center}
\includegraphics[width=10.5cm]{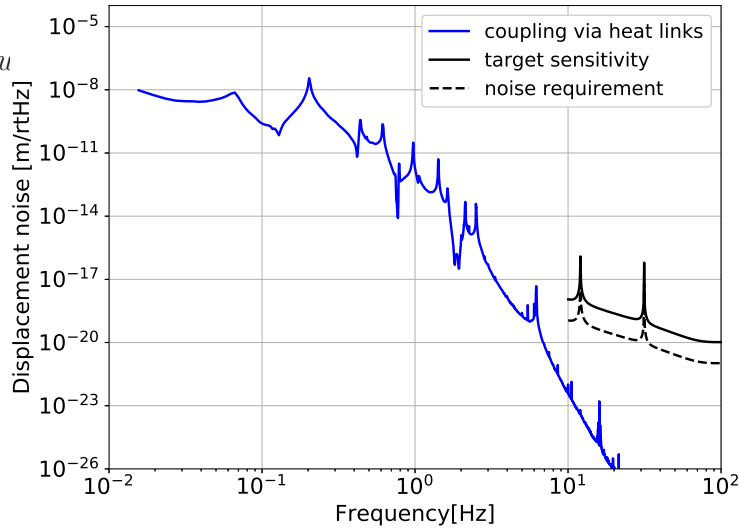}
\end{center}
\caption{Estimation of vibration noise via heat link. we use following parameters for calculation: $m=1.3\times10^{-2}$\,kg, which is corresponding to 0.8\,m heat link, $g=9.8$\,${\rm m/s^2}$, $\rho=0.0164$\,kg/m, and $x_0=0.545$\,m.}
\label{HLN}
\end{figure}

Here, we assume the attached heat link form a catenary, which is well known as hanging curve under gravity. Then, lateral force to the suspension due to the heat link satisfies following equation:
\begin{equation}
\frac{mg}{2T} = \sinh\frac{\rho gx}{2T},
\end{equation} 
where $m$, $g$, $\rho$ $x$, $T$ are mass of the heat links, gravitational acceleration, linear mass density of the heat link, distance between both ends of the heat link, and lateral force to the suspension, respectively. Considering small fluctuation of $x$ from the nominal distance $x_0$ due to vibration of cooling bar, fluctuation of $T$ from the nominal force $T_0$ can be calculated by following equation:
\begin{equation}
S_T(f) = \frac{\rho gT_0\cosh\left(\frac{\rho g x_0}{2T_0}\right)}{-mg+\rho g x_0\cosh\left(\frac{\rho g x_0}{2T_0}\right)} S_x(f),
\end{equation} 
where $S_T(f)$ and $S_x(f)$ are amplitude spectra of $T$ and $x$ as a function of Fourier frequency $f$, respectively. Considering the seismic spectrum as cooling bar vibration and using the rigid body model of the cryogenic payload \cite{SUMCON}, displacement noise coupling via heat link can be estimated. Figure \ref{HLN} shows the coupling of vibration through heat links. We note that vibration inside the cryostat might be larger than seismic noise at KAGRA site due to resonances of the cryostat and cryocooler vibration. This should be investigated in near future.

\begin{figure}
\begin{center}
\includegraphics[width=10.5cm]{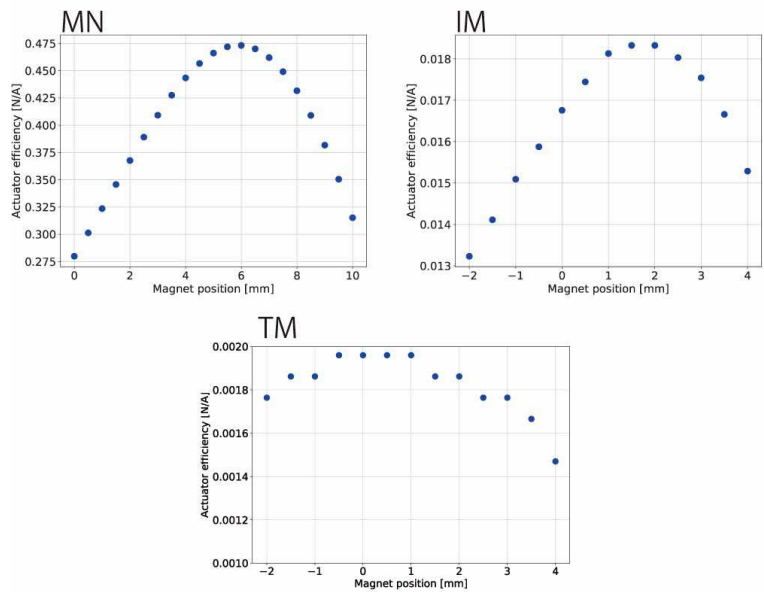}
\end{center}
\caption{Result of actuator efficiency measurement at room temperature. Magnet position is measured from the surface of the coil. Nominal magnet position of each stage is designed as the gradient of actuator efficiency is zero. (Left top) Actuator efficiency on the MN stage actuator. Actuator efficiency on the MN stage is 470\,mN/A. (Right top) Actuator efficiency on the IM stage actuator. Actuator efficiency on the IM stage is 18\,mN/A. (Bottom) Actuator efficiency on the TM stage actuator. Actuator efficiency on the TM stage is 2.0\,mN/A. }
\label{AEtest}
\end{figure}

PF is also used for adjusting the inclination of an RM chain to a TM chain by utilizing moving masses. Relative position of TM chain and RM chain is important to avoid touching each other after cooling because thermal contraction may cause large misalignment. In addition, it is also important to set magnets, which are located on MN, IM, and TM, at the position where magnetic-field gradient of the coil is zero for minimizing efficiency variation due to relative position change between the coil and magnet by thermal contraction. Figure \ref{AEtest} shows the result of coil-magnet actuator efficiencies when the relative position between coils and magnets are changed. Designed actuator efficiencies on MN, IM, and TM stages are 430\,mN/A, 16\,mN/A, and 1.5\,mN/A at cryogenic temperature \cite{MAP}, respectively. Measured actuator efficiencies on MN, IM, and TM stages at room temperature are 470\,mN/A, 18\,mN/A, and 2.0\,mN/A, respectively. Though the measured values are 10-30\% higher than designed values, they are acceptable. One reason is that the actuator efficiency will become lower when cooling because of reduction of magnetism of SmCo magnets. The other reason is that these differences can be compensated by modifying the driving circuit of coil-magnet actuators.

\subsection{Marionette stage}
\begin{figure}
\begin{center}
\includegraphics[width=10cm]{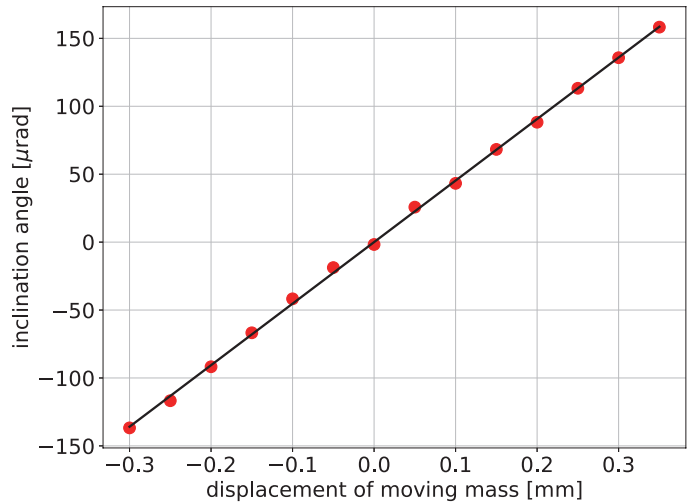}
\end{center}
\caption{Result of moving mass performance test. Change of angle is measured by an optical lever. Red dots and black line are measured vales and fitting line, respectively. The slope of fitting line is about 0.5\,rad/m.}
\label{MovingMassResult}
\end{figure}

An MN stage is used to adjust an inclination of TM to an injected beam. MN has two moving masses for rough alignment of a TM chain. Figure \ref{MovingMassResult} shows inclination changes of the sapphire mirror by using the moving mass system on MN. From this result, we can extrapolate the actuation range as $\pm$ 11\,mrad and evaluate a resolution as 0.6\,$\mu$rad. Since the coil-magnet actuator on MN stage can actuate mirror angles by more than 10\,$\mu$rad, the 0.6\,${\rm \mu}$rad resolution of the moving mass stage is adequate to provide continuous control with the coil-magnet stage over a $\pm$11\,mrad range.

Another role of the MN stage is to damp yaw motion of the TM chain. Since the TM chain is suspended by a single wire from the PF, yaw motion of MN is easy to be excited and disturbs interferometer operation. On the other hand, an RM chain is suspended by three rods and its yaw motion is rigid compared to that of the TM chain. So, relative displacement signal of PSs can be used to damp the yaw motion of the TM chain. In addition, since the MN stage has an AS OpLev, its signal can be also used for the damping.

The other role of MN is to control longitudinal motion at low frequencies during operation of the interferometer. To reduce noise from actuation, IM and TM stages have relatively small magnets compared to those of MN. So, longitudinal motion control at low-frequencies, where a large dynamic range is required, is implemented by using MN-stage actuators.

\subsection{Intermediate mass stage}
The most significant role of the IM stage is to suspend sapphire mirrors in a manner that reduces suspension thermal noise as much as possible. Thus, sapphire fibers, which have high Q-factor at low temperature, are used for mirror suspension. Fiber length manufacturing tolerance is about 0.1\,mm while the fiber elasticity only results in tens of micrometer stretch. It is thus difficult to absorb fabrication errors by elastic deformation of the fibers. Therefore, four sapphire blade springs are installed on the IM. Since they are deformed over 1\,mm when adding the load of a mirror on them, fabrication errors of sapphire fibers can be absorbed by the sapphire blade springs.

\begin{figure}
\begin{center}
\includegraphics[width=10cm]{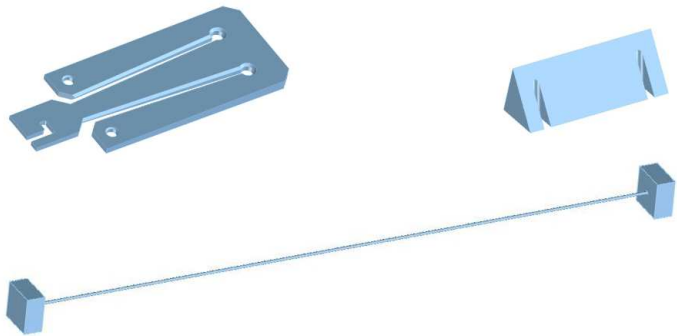}
\end{center}
\caption{(Top left) Schematic view of a sapphire blade spring. The slit for hooking the sapphire fiber is L-shape to prevent the fiber from slipping off. Sapphire blades have a taper to be compact and strong enough to suspend mirror. Thinner side (L-shape slit part) is 2\,mm. Thicker part (opposite L) is 4.5\,mm. (Top right) Schematic view of a sapphire ear. The ear shape is triangular prism 80\,mm in height. There are two slits 60\,mm apart from each other for hooking the sapphire fibers. (Bottom) Schematic view of a sapphire fiber. There are 20\,mm$\times$20\,mm$\times$10\,mm rectangular blocks at both ends of the fiber. The length and diameter of the fiber are 350\,mm and 1.6\,mm, respectively.}
\label{Fig2}
\end{figure}

The sapphire blade spring has an M-shape for obtaining low resonant frequency of bounce mode in a small space. The thickness of the blade is tapered to be strong enough to support the TM load, as described in Fig. \ref{Fig2}. The sapphire blade springs are attached to the IM inclined at an angle of 1.9 degrees (as are the CuBe blade springs attached to the PF). A schematic view of the sapphire blade spring, ear and fiber are shown in Fig. \ref{Fig2}.

The other role of the IM stage is to provide redundancy of damping eigenmodes of the suspension. Since our photosensor was designed to obtain large rage for accommodating thermal contraction during cooling, the sensitivity of the PS is lower than that of sensors used at room temperature suspension. It might cause difficulties to damp some eigenmodes effectively by using PSs on MN stage. Since the differential motion at the IM stage is larger than that at the MN stage in some eigenmodes, PSs on IM stage can be used for damping such modes in case that it is difficult to damp effectively at MN stage.

\subsection{Test mass stage}
The TM is the most important part of the cryogenic payload for achieving design sensitivity of KAGRA \cite{PSO}. The TM is made of monocrystalline sapphire that has high mechanical Q-factor, high thermal conductivity \cite{STC}, and low absorption of 1064\,nm light (50\,ppm/cm) \cite{OptAbs}. The TM also has high reflective or anti-reflective multi-layer coatings for 1064\,nm light, on the front and back circular surfaces of the cylinder respectively. These coating also have low, sub ppm optical absorption \cite{CoatAbs}.

\begin{figure}
\begin{center}
\includegraphics[width=7cm]{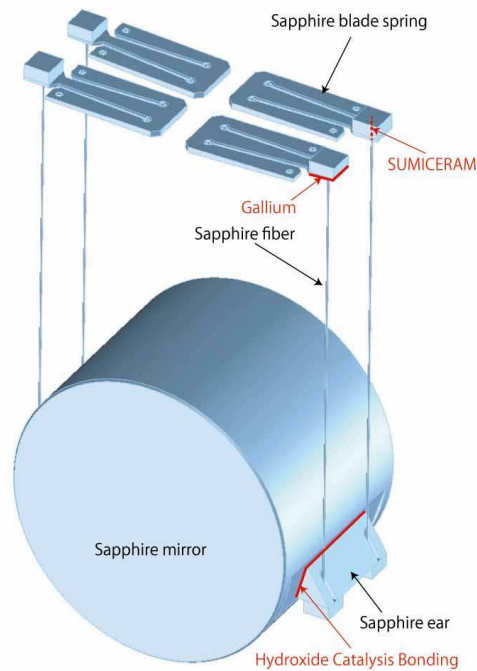}
\end{center}
\caption{Schematic view of the KAGRA sapphire suspension. A sapphire mirror has two ears, which is bonded by HCB, for hooking sapphire fibers. The mirror is suspended from the sapphire blade spring on IM to mitigate fiber length difference. Both edge of sapphire fiber has rectangular blocks, which is bonded by inorganic adhesive called SUMICERAM. These rectangular blocks are bonded to the ears and blade springs with Gallium in order not to slip off and to have sufficient thermal contact. Names of each parts and bodings are written in the figure with black and red letters, respectively.}
\label{fig_SS}
\end{figure}

To suspend mirrors, we use sapphire fibers with nail heads and sapphire ears with slits. Figure \ref{Fig2} shows a schematic view of a sapphire ear and sapphire fiber. The sapphire fiber has rectangular blocks at either end to hook onto the ear and blade spring. The blocks are bonded by an inorganic adhesive called SUMICERAM. The sapphire ear is a triangular prism with 80\,mm in height. The ear has two slits apart from 60\,mm each other to hook the sapphire fibers. Sapphire ears are bonded at the side of a sapphire mirror and suspended from sapphire blade spring on the IM as shown in Fig. \ref{fig_SS}

Bonding between the mirror and ear should be strong enough to support the mirror weight at cryogenic temperature. In addition, mechanical loss of bond must be small enough not to increase thermal noise. Since hydroxide catalysis bonding (HCB) method \cite{HCB} satisfies both high strength and low mechanical losses \cite{ST1_HCB,ST2_HCB,Q_HCB}, HCB is used for the bonding between the mirror and ears. Sapphire ears are attached at a position 47.8\,mm below the center of mass of the mirror. This is because the bending point of the fiber should be in the same horizontal plane as the center of mass to minimize coupling between TM pitch motion and longitudinal motion.

Bonding between the fiber nail head and the ears or blade is also required to prevent slip and to ensure good thermal contact. Moreover, it must be possible to replace fibers in case of fiber breakage or other accidents. We thus use Gallium (Ga) for bonding them. The melting point of Ga is 30 deg C, therefore it is easy to melt. Also, the bond can be easily separated by warming Ga. Since Ga cannot adhere to sapphire surface tightly by just melting it on sapphire, it is applied to sapphire surfaces by using an ultrasonic soldering iron. This method prevents the Ga from peeling off. This bond is also used between the other nail head and sapphire blade spring. After application of Ga and suspending the mirror by four sapphire fibers, heat is applied to Ga for melting them and bond the ears and fibers. Figure \ref{fig_SS} shows the sapphire suspension system used in KAGRA.

\section{Thermal design of a cryogenic payload}%%%%%%%%%%%%%%%%%%%%%%%%%%%%%%%%%%%%%%%%%%%%%%%%%%%%%%%%%%%%%%%%%%%%%%%%%%%%%
\subsection{Overview}
A cryogenic payload is cooled by thermal radiation and heat conduction in high and low temperature region, respectively. For effective radiation cooling, the cryogenic payload is coated by a black plating called Solblack coating that has a high emissivity. Four cryocoolers are used to maintain an outer radiation shield at 80\,K. Two cryocoolers have a second stage to maintain an inner radiation shield at 8\,K for radiative cooling, while two other cryocoolers' second stage is dedicated to conductive cooling of the payload. Soft heat links are connected between the cryogenic payload and two cooling bars as shown in Fig. \ref{Type-A}, which have a good thermal connection to the second stage of cryocoolers.

There are two layers of radiation shields around the cryogenic payload for reducing thermal radiation from the room temperature environment. Both radiation shields have multi-layers insulation on their outer surfaces. The insulation is made of polyimide films with evaporated aluminum for reducing the heat load by thermal radiation. The inner surface of both radiation shields are coated by a diamond like carbon coating to make heat extraction via thermal radiation work effectively. Cryogenic baffles are also installed in the beam duct of the main interferometer to reduce thermal radiation.

Heat links for cooling in low temperature region are made of pure aluminum, which has a purity of over 99.9998\% \cite{HL}. This ultra pure aluminum heat link has high thermal conductivity, to cool the cryogenic payload effectively, and flexibility to reduce the vibration coupling through the heat links. Four heat links, 800\,mm in length, are first connected from cooling bars to MNR for reducing the vibration coupling to the TM.

The cryogenic payload is cooled to 20\,K by two double-stage pulse-tube cryocoolers (PTCs) that create very small vibration \cite{PTC, CS}. The inner radiation shield is also cooled to 8\,K by two other PTCs. The outer radiation shield is cooled by the first-stage of all four PTCs that are used for cooling of the cryogenic payload and inner radiation shield.

Beam ducts on both the HR side and AR side have cryogenic duct-shields 5\,m in length which are installed to reduce radiative heat load from the room-temperature beam duct. They also have Solblack coating to obtain effective reduction of thermal radiation and are cooled to 120\,K by a single-stage PTC. Thanks to these cryogenic duct shields, thermal radiation is reduced by a factor of 100 resulting in 0.1\,W heat load per beam duct \cite{CDS}.

\subsection{Cryogenic payload}
The KAGRA mirror obtains heat of about 0.7\,W by absorbing the main laser of an interferometer when operating at full power of KAGRA design even though mirror substrate and coatings have very low absorption of 1064\,nm light. In order to extract such large heat without any large mechanical losses, sapphire fibers, which have a Q-factor of the order of $10^7$, of a diameter of 1.6\,mm with nail heads are used for suspension. HCB and bond of Ga are also essential for reducing the thermal resistance of the sapphire suspension.

Mirror thermal noise and suspension thermal noise have been calculated and shown in Ref \cite{bKAGRA-phase1}. Thanks to the cryogenic mirrors, mirror thermal noise is not a dominant noise source in the configuration of broadband resonant side-band extraction (RSE), while suspension thermal noise is a dominant noise source at low frequencies because sapphire fibers for suspending mirrors are thick to extract the heat from the mirrors. In the case of detuned RSE, both mirror and suspension thermal noise are dominant noise sources at middle and low frequencies, respectively. To achieve the KAGRA target sensitivity, KAGRA mirror needs to be cooled down below 22\,K when injecting laser power of 67\,W to the power recycling cavity. We note that we assume the effect of bonding can be negligible in that calculation. Thus, further investigations on thickness and Q-factor of bonding are necessary to confirm that bonding doesn't contaminate thermal noise of KAGRA.

\begin{figure}
\begin{center}F
\includegraphics[width=14cm]{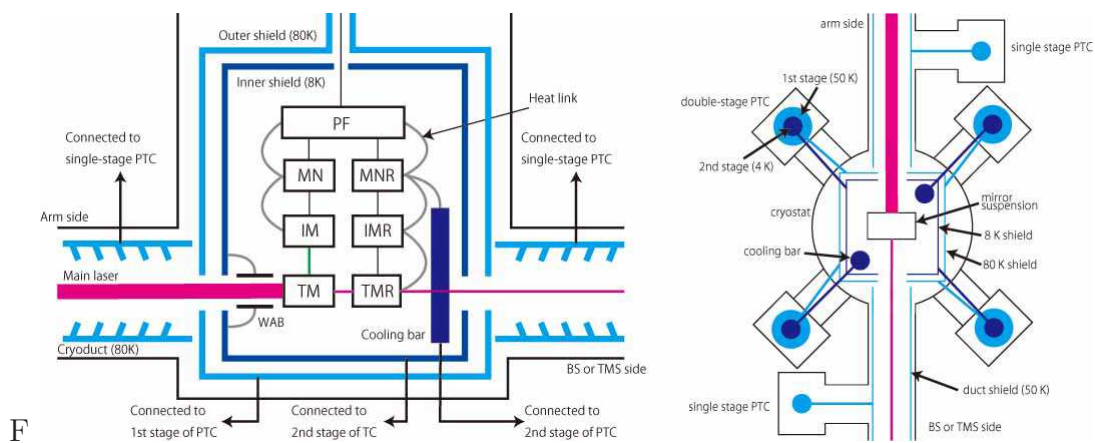}
\end{center}
\caption{(Left) Schematic view of the KAGRA cryogenic system seen from the side. Heat extraction from sapphire mirror is through 4 sapphire fibers, and the other part of payload is cooled by using pure-aluminum heat links. MNR is connected to cooling bars, which is directly contacted to the second stage of 4\,K cryocooler, by the same heat links. A wide angle baffle \cite{WAB}, which is for absorbing scattered light and results the reduction of scattered light noise, is placed on the inner radiation shield and the heat links are connected for heat extraction. Cryogenic duct-shield, which is cooled to 120\,K by single-stage cryocooler, located at the both sides of cryostat along the beam tube. (Right) Schematic view seen from the top side. outer radiation shield is cooled by 4 first stage of two-stage PTCs, while inner shield is cooled by 2 second stage of them.}
\label{CoolingSystem}
\end{figure}

The heat is transmitted in order of IM, MN, PF, and MNR after transferring through the sapphire suspension. RM are also cooled and the heat is transmitted in order of IRM and MNR through aluminum heat links. MNR is linked to cooling bars that are directly cooled by the cryocoolers. A schematic diagram of the cooling system is shown in Fig. \ref{CoolingSystem}.

The first KAGRA cryogenic payload was installed at the end of November, 2017 and was cooled down from the beginning of February, 2018. There were thermometers on the inner and outer radiation shields, cooling bars, and each stage of the payload: TM, RM, IM, IRM, MN, MNR, and PF.

\begin{figure}
\begin{center}
\includegraphics[width=14cm]{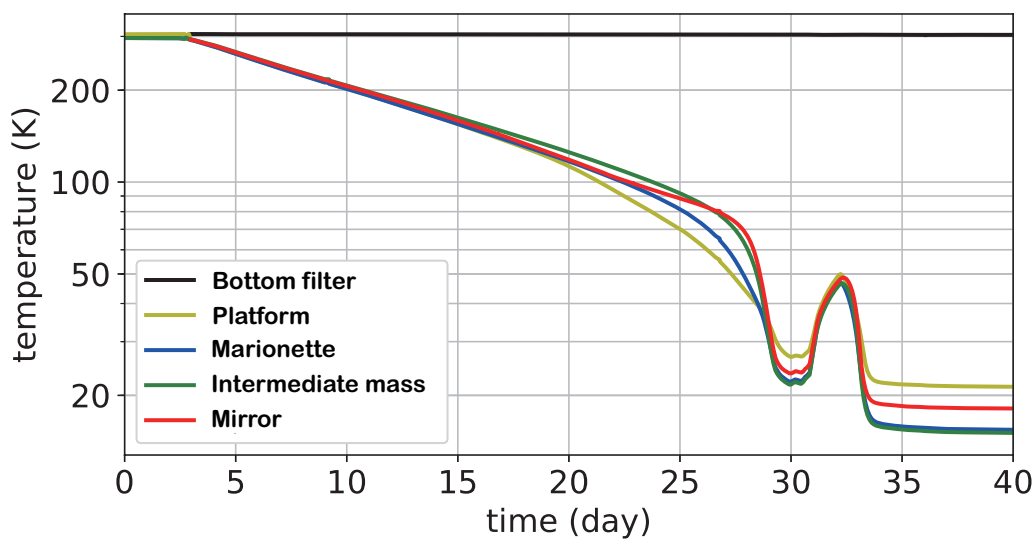}
\end{center}
\caption{Cooling curve during bKAGRA Phase 1. Figure is adopted and revised from Ref. \cite{bKAGRA-phase1}. Red curve, showing sapphire mirror temperature, reaches below 20\,K within one month. Black line shows the temperature of BF, which is the nearest stage from the cryogenic payload. Even though the temperature of cryogenic payloads goes down, BF can be kept at room temperature because of the well-thermally-isolated system.}
\label{CoolingTime}
\end{figure}

Figure \ref{CoolingTime} shows the temperature of the suspension during cooling. Unfortunately, we couldn't measure temperature of MNR because its thermometer didn't work at that time. It took almost one month to reach the steady state mirror temperature of about 18\,K. This is approximately consistent with our estimation of cooling time and is acceptable for KAGRA operation. From 27th day to 29th day, temperature change of the payload was stopped because the temperature of one of the cryocoolers for cooling the payload stops at 23\,K while the temperature of the other reached 10\,K. Thus, we once stopped cryocoolers on 29th day and restarted on 30th day. Then, the mirror temperature reached steady state at 34 days. We note that this situation doesn't seem to come from the trouble of that particular cryocooler because we have faced the similar situation several times.

\section{Discussion and Conclusion}%%%%%%%%%%%%%%%%%%%%%%%%%%%%%%%%%%%%%%%%%%%%%%%%%%%%%%%%%%%%%%%%%%%%%%%%%%%%%%%%%%%%%%%%
During cooling of the cryogenic payload, angular drift of pitch and yaw were observed. the total drifts during cooling were about 100\,${\rm \mu rad}$ and hundreds of ${\rm \mu rad}$ for pitch and yaw, respectively. For pitch drift compensation, the moving mass on MN stage was used while BF actuators were used for yaw. Then, the mirror can be successfully aligned by using these actuators for operation \cite{bKAGRA-phase1}. After the alignment, commissioning for a simple Michelson interferometer was performed. 

A test operation, bKAGRA Phase 1, was performed for 10 days after the commissioning. During bKAGRA Phase 1, transfer functions from voltage applied to the coil drivers to the displacement of the mirror was measured \cite{bKAGRA-phase1}. Actuator efficiency on TM stage can be calculated as 1\,mN/A from the result around 10\,Hz, which is half of the actuator efficiency measured at room temperature. The measured actuator efficiency is two thirds smaller than our designed values but it can be acceptable because the difference can be compensated by modifying the driving circuit of coil-magnet actuators. We note that actuator efficiency of IM and MN stages are difficult to be evaluated from the result because there are large spurious couplings that could come from the issue of cables between BF and PF.

During bKAGRA Phase 1, we identified three issues. The first one is that the electrical cables running between BF and PF through a narrow cylindrical radiation aperture were found to be rubbing against the aperture, introducing large amount of vibration to TM. The issue was resolved by tying the cables to the suspension rod without modifying the cylindrical aperture. This preserved the radiation shielding between the room temperature upper parts and the cryogenic payload. Details are reported in Ref. \cite{bKAGRA-phase1}. The second issue is that one of yaw resonant modes, which twists the whole cryogenic payload commonly, was found to be difficult to damp by using the existing sensors and actuators of Type-A suspension. Therefore, the MN actuators and AS OpLev were used to control the mirror angle with respect to the ground and to reduce the angular motion of the mirror during bKAGRA Phase 1. To resolve this issue, coil-magnet actuators and OpLevs are planned to be installed on the PF stage. The final issue is long-term reliability of moving masses. The moving mass can work at cryogenic temperature and has a good performance during the initial alignment. However, the ball screws lose their smooth operation when passing thermal cycles, and then some moving masses don't work after warming up. The reason could be friction between balls inside the ball screws. Since ball screws are oil-free, friction between balls creates metal dusts inside the ball screws and thus the ball screws can be stuck. Therefore, a new moving mass system, which doesn't use ball screws, was designed to solve this problem. Pulley and thin wire is used for driving a copper mass instead of a ball screw in the new design. A prototype test is currently underway.

We designed and demonstrated the first cryogenic suspension for a km-scale interferometric gravitational-wave detector. A 23\,kg sapphire mirror was suspended by four sapphire fibers 1.6\,mm in diameter and cooled below 20\,K by conduction cooling via pure aluminum heat links and the cooling time is almost the same as what we expected. Vibration coupling via heat link is estimated and the sensors and actuators work well at cryogenic temperature and show sufficient performances for the operation of the 3-km interferometer. No fatal issues were found and several issues found in the first cryogenic payload have been resolved. We also survey several upgrades based on our first design of the cryogenic payload \cite{UPG}. This design paves the way to LIGO Voyager, Cosmic Explorer, and Einstein Telescope \cite{Voyager-CE,ET}.

\section*{Acknowledgement}%%%%%%%%%%%%%%%%%%%%%%%%%%%%%%%%%%%%%%%%%%%%%%%%%%%%%%%%%%%%%%%%%%%%%%%%%%%%%%%%%%%%%%%%%%%%%%
This work was supported by MEXT, JSPS Leading-edge Research Infrastructure Program, JSPS Grant-in-Aid for Specially Promoted Research 26000005, JSPS Grant-in-Aid for Scientific Research on Innovative Areas 2905: JP17H06358, JP17H06361 and JP17H06364, JSPS Core-to-Core Program A. Advanced Research Networks, JSPS Grant-in-Aid for Scientific Research (S) 17H06133, the joint research program of the Institute for Cosmic Ray Research, University of Tokyo, National Research Foundation (NRF) and Computing Infrastructure Project of KISTI-GSDC in Korea, Academia Sinica (AS), AS Grid Center (ASGC) and the Ministry of Science and Technology (MoST) in Taiwan under grants including AS-CDA-105-M06, the LIGO project, and the Virgo project. We thank the Advanced Technology Center (ATC) of NAOJ and the Engineering Machine Shop, Faculty of Engineering, University of Toyama for supporting development of sapphire suspensions. We also thank the Mechanical Engineering Center of KEK for fabricating many items of the cryogenic payloads.


\section*{References}
\begin{thebibliography}{99}
\bibitem{GW150914} B. P. Abbott {\it et al.}, Observation of Gravitational Waves from a Binary Black Hole Merger. 2016, Phys. Rev. Lett., \href{https://link.aps.org/doi/10.1103/PhysRevLett.116.241103}{{\bf 116} 061102}
\bibitem{GWTC-1} B. P. Abbott {\it et al.}, GWTC-1: A Gravitational-Wave Transient Catalog of Compact Binary Mergers Observed by LIGO and Virgo during the First and Second Observing Runs. 2019, Phys. Rev. X., \href{https://link.aps.org/doi/10.1103/PhysRevX.9.031040}{{\bf 9} 031040}
\bibitem{GRT150914} B. P. Abbott {\it et al.}, 2016, Tests of General Relativity with GW150914. Phys. Rev. Lett.., \href{https://link.aps.org/doi/10.1103/PhysRevLett.116.221101}{{\bf 116} 221101}
\bibitem{BHBHCRE1} B. P. Abbott {\it et al.}, Binary Black Hole Population Properties Inferred from the First and Second Observing Runs of Advanced LIGO and Advanced Virgo. 2019, The Astrophysical Journal, \href{https://doi.org/10.3847%2F2041-8213%2Fab3800}{{\bf 882} L24}
\bibitem{NSNSCRE1} B. P. Abbott {\it et al.}, UPPER LIMITS ON THE RATES OF BINARY NEUTRON STAR AND NEUTRON STAR-BLACK HOLE MERGERS FROM ADVANCED LIGO'S FIRST OBSERVING RUN. 2016, The Astrophysical Journal, \href{https://doi.org/10.3847%2F2041-8205%2F832%2F2%2Fl21}{{\bf 832} L21}
\bibitem{BBHMO1} B. P. Abbott {\it et al.}, Properties of the Binary Black Hole Merger GW150914. 2016, Phys. Rev. Lett.., \href{https://link.aps.org/doi/10.1103/PhysRevLett.116.241102}{{\bf 116} 221102}
\bibitem{HCM170817} LIGO Scientific Collaboration, Virgo Collaboration, 1M2H Collaboration, Dark Energy Camera GW-EM Collaboration, DES Collaboration, DLT40 Collaboration, Las Cumbres Observatory Collaboration, VINROUGE Collaboration, and MASTER Collaboration, A gravitational-wave standard siren measurement of the Hubble constant. 2017, Nature., \href{https://doi.org/10.1038/nature24471}{{\bf 551} 85-88}
\bibitem{LIGO} J. Aasi {\it et al.}, Advanced LIGO. Class. Quant. Grav.., \href{https://doi.org/10.1088/0264-9381/32/7/074001}{{\bf 32} 074001}
\bibitem{Virgo} F. Acernese {\it et al.}, Advanced Virgo: A second-generation interferometric gravitational wave detector. 2015, Class. Quant. Grav.., \href{https://doi.org/10.1088/0264-9381/32/2/024001}{{\bf 32} 024001}
\bibitem{KAGRA} Yoichi Aso {\it et al.}, Interferometer design of the KAGRA gravitational wave detector. 2013 Phys. Rev. D., \href{https://doi.org/10.1103/PhysRevD.88.043007}{{\bf 88} 1-15}
\bibitem{sapphireQ} T. Uchiyama {\it et al.}, Mechanical quality factor of a cryogenic sapphire test mass for gravitational wave detectors. 1999, Phys. Rev. A., \href{https://www.sciencedirect.com/science/article/pii/S0375960199005630}{{\bf 261} 5-11}
\bibitem{bKAGRA-phase1} T. Akutsu {\it et al.}, First cryogenic test operation of underground km-scale gravitational-wave observatory {KAGRA}. 2019, Class. Quant. Grav.., \href{https://doi.org/10.1088/1361-6382/ab28a9}{{\bf 36} 165008}
\bibitem{MAP} Yuta Michimura {\it et al.}, Mirror actuation design for the interferometer control of the KAGRA gravitational wave telescope. 2017, Class. Quant. Grav.., \href{https://doi.org/10.1088/1361-6382/aa90e3}{{\bf 34} 225001}
\bibitem{GAS} G. Cella {\it et al.}, Seismic attenuation performance of the first prototype of a geometric anti-spring filter. 2002, Nucl. Instrum. Meth. Phys. Res. A., \href{https://doi.org/10.1016/S0168-9002(01)02193-3}{{\bf 487} 652-660}
\bibitem{SUMCON} SUMCON (available from JGW-\href{https://gwdoc.icrr.u-tokyo.ac.jp/cgi-bin/DocDB/ShowDocument?docid=3729}{T1503729})
\bibitem{oplev} S. Zeidler, Length-Sensing OpLevs for KAGRA. JGW - \href{https://gwdoc.icrr.u-tokyo.ac.jp/cgi-bin/private/DocDB/ShowDocument?docid=5788}{T1605788}
\bibitem{PS} T. Ushiba {\it et al.}, preparation
\bibitem{OSEM} L. Carbone {\it et al.}, Sensors and actuators for the Advanced LIGO mirror suspensions. 2012, Class. Quant. Grav.., \href{https://doi.org/10.1088/0264-9381/29/11/115005}{{\bf 29} 115005}
\bibitem{OSEMK} T. Akutsu {\it et al.}, Compact integrated optical sensors and electromagnetic actuators for vibration isolation systems in the gravitational-wave detector KAGRA, Rev. Sci. Instrum., \href{https://doi.org/10.1063/5.0022242}{{\bf 91} 115001}
\bibitem{magnet} Stanley R. Trout {\it et al.}, Using Permanent Magnets at Low Temperatures. 2005, \href{http://spontaneousmaterials.com/Papers/TN_0302.pdf}{http://spontaneousmaterials.com/Papers/TN\_0302.pdf}
\bibitem{PSO} Yuta Michimura {\it et al.}, Particle swarm optimization of the sensitivity of a cryogenic gravitational wave detector, 2018, Phys. Rev. D, \href{https://doi.org/10.1103/PhysRevD.97.122003}{{\bf 97} 122003}
\bibitem{STC} A. Khalaidovski {\it et al.}, Evaluation of heat extraction through sapphire fibers for the GW observatory KAGRA. 2014, Class. Quant. Grav., \href{https://doi.org/10.1088/0264-9381/31/10/105004}{{\bf 31} 105004}
\bibitem{OptAbs} Takayuki. Tomaru {\it et al.}, Cryogenic measurement of the optical absorption coefficient in sapphire crystals at 1.064 $\mu$m for the large-scale cryogenic gravitational wave telescope. 2001, Phys. Rev. A., \href{https://doi.org/10.1016/S0375-9601(01)00191-8}{{\bf 283} 80-84}
\bibitem{CoatAbs} E. Hirose, Charcterization of the coated ITMs. JGW - \href{https://gwdoc.icrr.u-tokyo.ac.jp/cgi-bin/private/DocDB/ShowDocument?docid=9173}{T1809173}
\bibitem{HCB} E. J. Elliffe {\it et al.}, Hydroxide-catalysis bonding for stable optical systems for space. 2005, Class. Quant. Grav.., \href{https://doi.org/10.1088/0264-9381/22/10/018}{{\bf 22} S257-S267}
\bibitem{ST1_HCB} K. Haughian {\it et al.}, The effect of crystal orientation on the cryogenic strength of hydroxide catalysis bonded sapphire. 2015, Class. Quant. Grav., \href{https://doi.org/10.1088/0264-9381/32/7/075013}{{\bf 32} 075013}
\bibitem{ST2_HCB} Margot Phelps {\it et al.}, Strength of hydroxide catalysis bonds between sapphire, silicon, and fused silica as a function of time. 2018, Phys. Rev. D., \href{https://doi.org/10.1103/PhysRevD.98.122003}{{\bf 98} 122003}
\bibitem{Q_HCB} K. Haughian {\it et al.}, Mechanical loss of a hydroxide catalysis bond between sapphire substrates and its effect on the sensitivity of future gravitational wave detectors. 2016, Phys. Rev. D, \href{https://doi.org/10.1103/PhysRevD.94.082003}{{\bf 94} 082003}
\bibitem{HL} T. Yamada {\it et al.}, High Performance Heat Conductor with Small Spring Constant for Cryogenic Applications. arXiv: \href{https://arxiv.org/abs/2003.13457}{2003.13457}
\bibitem{PTC} Y. Ikushima {\it et al.}, Ultra-low-vibration pulse-tube cryocooler system - cooling capacity and vibration. Cryogenics., \href{https://doi.org/10.1016/j.cryogenics.2008.04.001}{{\bf 48} 406-412}
\bibitem{CS} C. Tokoku {\it et al.}, Cryogenic system for the interferometric cryogenic gravitationalwave telescope, KAGRA - design, fabrication, and performance test -. 2014, AIP Conference Proceedings, \href{https://doi.org/10.1063/1.4860850}{{\bf 1573} 1254}
\bibitem{CDS} Y. Sakakibara {\it et al.}, Performance test of pipe-shaped radiation shields for cryogenic interferometric gravitational wave detectors., 2015, Class. Quant. Grav., \href{https://doi.org/10.1088/0264-9381/32/15/155011}{{\bf 32} 155011}
\bibitem{WAB} T. Akutsu {\it et al.}, Vacuum and cryogenic compatible black surface for large optical baffles in advanced gravitational-wave telescopes, 2016, Opt. Mater. Express, \href{https://doi.org/10.1364/OME.6.001613}{{\bf 6} pp. 1613-1626}
\bibitem{UPG} Y. Michimura {\it et al.}, Prospects for improving the sensitivity of KAGRA gravitational wave detector.  arXiv: \href{https://arxiv.org/abs/1906.02866}{1906.02866}
\bibitem{Voyager-CE} LIGO Scientific Collaboration, Instrument Science White Paper 2019. 2019, LIGO-\href{https://dcc.ligo.org/LIGO-T1900409-v4/public}{T1900409}
\bibitem{ET} M Punturo {\it et al.}, The Einstein Telescope: a third-generation gravitational wave observatory. 2010, Class. Quantum Grav. \href{https://doi.org/10.1088/0264-9381/27/19/194002}{{\bf 27}, 194002}
\end{thebibliography}
\end{document}